\documentstyle[latexsym]{article}

\def\be{\begin{eqnarray}}
\def\ee{\end{eqnarray}}

\hoffset= - 1.0in         

\title{{\bf Is Strong Gravitational Radiation predicted by TeV-Gravity?
} \vspace{.5cm}}
\author{{\bf A. Mironov}\footnote{E-mail: \ mironov@lpi.ru; mironov@itep.ru}
\date{ } \\
{\small {\it Lebedev Physics Institute}
and {\it ITEP, Moscow, Russia}}\\ \\
{\bf A.Morozov}\thanks{E-mail: \ morozov@itep.ru}
\date{ } \\ {\small {\it ITEP, Moscow, Russia}}
}

\begin{document}

\maketitle

\vspace{-8.0cm}

\begin{center}
\hfill FIAN/TD-11/06\\
\hfill ITEP/TH-64/06\\
\end{center}

\vspace{6.5cm}

\begin{abstract}
\noindent In TeV-gravity models 
the gravitational coupling to particles with energies
${\cal E} \sim m_{Pl} \sim 10\ TeV$
is {\it not} suppressed by powers of ultra-small ratio
${\cal E}/M_{Pl}$ with $M_{Pl} \sim 10^{19}GeV$.
Therefore one could imagine strong synchrotron
radiation of gravitons by the accelerating particles to 
become the
most pronounced manifestation of TeV-gravity at LHC.
However, this turns out to be not true:
considerable damping continues to exist, only the place
of ${\cal E}/M_{Pl}$ it taken by
a power of a ratio $\vartheta\omega/{\cal E}$, where
the typical frequency $\omega$ of emitted radiation,
while increased by a number of $\gamma$-factors,
can not reach ${\cal E}/\vartheta$
unless particles are accelerated by nearly critical fields.
Moreover, for currently available magnetic fields
$B \sim 10\ {\rm Tesla}$,
multi-dimensionality does not enhance gravitational
radiation at all even if TeV-gravity is correct.
\end{abstract}

\bigskip

Observation of gravitational waves is one of the most
challenging problems for modern experimental physics.
Most efforts have been \cite{astrograpast}
and continue to be \cite{astrogranew} concentrated
on the search of such waves of astrophysical origin,
and justified hopes exist that they will be indeed found
by the new generation of gravitational detectors.
Such detectors are made from superheavy bodies,
slightly excited by passing long-wave excitations of
gravitational background.
Of course,
of much more interest would be earthly
Hertz-style experiments, where relatively-short
gravitational waves are both generated and captured by
human-made devices.
Remarkably such ultra-high-precision experiments
are not as fantastic as they
can seem, and interesting ideas are already discussed
\cite{gr,chiao}
even in the framework of ordinary general relativity.

String-inspired TeV-gravity models \cite{TEVG}
provide an additional stimulus for the study of
various gravitational effects in four and higher
space-time dimensions,
because they can be potentially observable at LHC and
other high-precision experiments in the near future.
In this paper we address the issue of
ultra-short-wave gravitational radiation in TeV-gravity.
Surprisingly, the issue of radiation beyond four
dimensions is poorly represented in the literature and
incomplete results of ref.\cite{mudirad,koch}
turn insufficient for our purposes.
For systematic radiation theory,
generalizing \cite{4drad} to multidimensional situation,
see a separate paper \cite{mudirmm}.

In compactified multi-dimensional theories
gravitational interaction can be much stronger
than it is in four dimensions: the huge
Plank mass $M_{Pl}\sim 10^{19}GeV$ in four dimensions
can be made from a much smaller $4+n$-dimensional mass
$m_{Pl}$ and a large size $r_{KK} = m_{KK}^{-1}$
of $n$ compactified dimensions: 
\be
M_{Pl}^2 = \frac{m_{Pl}^{2+n}}{m_{KK}^n} =
m_{Pl}^2\left(\frac{m_{Pl}}{m_{KK}}\right)^n
\label{Mmrel}
\ee
If the last ratio $\frac{m_{Pl}}{m_{KK}}$ is big,
the ratio $\frac{M_{Pl}}{m_{Pl}}$ can be also big.
The value of $r_{KK}$ is severely restricted from above
by available experimental data
in ordinary Kaluza-Klein (KK) models, where all sorts of matter
are allowed to propagate in $4+n$-dimensional space-time,
but these restrictions are very weak in TeV-Gravity
models, where only gravity can propagate in the bulk,
while all other fields are confined within the $4d$ world-volume
of a $3$-brane. Actually, the values of $m_{Pl}$ as small
as TeV are not yet experimentally forbidden and
this opens a possibility of observing strong gravitational
effects already at the LHC.

\subsection{Gravitational effects at LHC predicted by TeV-gravity
models}

For particles with energies $E\sim m_{Pl}$ gravitational
interactions become as important as all other interactions,
provided momenta and energy transfers are also of the order
of $m_{Pl}$.
Thus gravitational effects can substantially change
cross sections and provide new types of events,
associated with exchange and flow-away of high-energy
gravitons.

Much more interesting can be strong classical
gravitational effects, like creation of
mini-black-holes \cite{mbh},
mini-black-rings \cite{mbr} and
mini-time-machines \cite{mtm}.
Elementary estimates imply that LHC can become a
factory, producing these kinds of so far exotic objects
at the rate of one-per-second, provided accelerator
energy be around $m_{Pl}$.
Actually all these objects would immediately evaporate
due to the Hawking radiation ( for the typical time of $10^{-28}$ s)
so that they behave like a
short-living ball of hadronic matter and their traces can be rather
hard to identify in experimental data despite the high
production rates.

It can seem that a much more pronounced effect can be
associated with gravitational radiation.
Indeed, in particle (especially electron) accelerators electromagnetic
radiation is not just observable:
it is tremendously large and rather difficult to exclude.
One could expect that whenever gravity gets strong,
the same happens with the gravitational radiation,
only -- in variance with the electromagnetic one -- it would
be of greatest interest for today's science.
The purpose of this paper is to explore this obvious
idea, and -- not too much surprisingly -- the answer turns
to be far more pessimistic: strong gravity does {\it not}
immediately imply strong gravitational radiation,
and even if TeV-gravity happens to be true but the
number of extra dimensions $n>2$, the
gravitational radiation in most realistic processes
will remain weak and hard to detect, may be even harder
than the mini-black-hole effects.

\subsection{In what sense is TeV-Gravity Strong?}

The usual way to derive important relation (\ref{Mmrel}),
justifying a possible enhancement of the gravitational
interaction in multidimensional theories, at the level
of Einstein-Hilbert action is
\be
m_{Pl}^{n+2}\int \int R^{(4+n)} d^4x d^ny
\ 
\longrightarrow\
m_{Pl}^{n+2}V_n \int R^{(4)} d^4x \sim
M_{Pl}^2 \int R^{(4)} d^4x
\ee
where the multi-dimensional metric is substituted in the
factorized form,
$g^{(4+n)}(x,y) \rightarrow g^{(4)}(x)\otimes g^{(n)}(y)$,
and the volume $V_n$ of compactified  dimensions is
proportional to $r_{KK}^n$ (for asymmetric compactifications,
when some dimensions are larger than others,
there are additional factors in
eq.(\ref{Mmrel}) which we ignore in
our approximate considerations).

Another derivation of (\ref{Mmrel}) can be done in terms of the
inverse-square Newton law.
One can easily see how the fast-falling
multi-dimensional potential of two-body interaction turns into a slower-falling
$4d$ one at expense of substantial decrease of the
coupling constant $m_{Pl}^{-1}\rightarrow M_{Pl}^{-1}$:
$$
\frac{1}{m_{Pl}^{(n+2)}}
\frac{m_1m_2}{\Big((\vec x-\vec x')^2 + (\vec y-\vec y')\Big)^{n+1\over 2}}
\ \stackrel{{\rm compactification}}{\longrightarrow}\
\frac{1}{m_{Pl}^{(n+2)}}\sum_{k_1\ldots \ k_n=-\infty}^\infty
\frac{m_1m_2}
{\Big((\vec x-\vec x')^2 + (\vec y-\vec y'+ \vec k \cdot
r_{KK})^2\Big)^{n+1\over 2}}
$$ \vspace{-0.2cm}
\be
\stackrel{|x-x'| \gg r_{KK}}{\longrightarrow}\
\frac{1}{m_{Pl}^{(n+2)}}\int {d^n y\over r_{KK}^n}
\frac{m_1m_2}
{\Big((\vec x-\vec x')^2 + {\vec y}^2\Big)^{n+1\over 2}}=
\frac{1}{m_{Pl}^{(n+2)}}\frac{1}{r_{KK}^n}\frac{m_1m_2}{|x-x'|}
= \frac{m_{KK}^n}{m_{Pl}^{(n+2)}}\frac{m_1m_2}{|x-x'|} =
\frac{1}{M_{Pl}^{2}}\frac{m_1m_2}{|x-x'|}
\ee
Here $\vec x$ and $\vec y$ refer to the uncompactified and compactified
coordinates accordingly, and the sum over $\vec k$ emerging due to
periodicity in compactified dimensions is replaced by the
integral, since $|\vec x-\vec x'|\gg r_{KK}$.

Reversing this argument is more interesting:
how do we get a big coupling constant $\sim m_{Pl}^{-1}$
at small distances starting from the small one
$\sim M_{Pl}^{-1}$ at large distances?
The point is that in the $4d$ theory, arising as
compactification from higher dimensions,
instead of a single graviton
one has a whole tower of KK excitations
labeled by integer-valued $n$-vector $\vec k$ with
masses $km_{KK}= k/r_{KK}$, $k = |\vec k|$ and the $4d$
two-body potentials have the Yukawa form
\be
\frac{m_1m_2}{M_{Pl}^2}\frac{e^{-k|x-x'|/r_{KK}}}{|\vec x-\vec x'|}
\ee
For $|x-x'|\ll r_{KK}$ they all contribute
up to $k = |\vec k| \sim \frac{r_{KK}}{|x-x'|}$,
which provides an enhancement factor $\sim k^n$
\be
\sum_{\vec k}
\frac{m_1m_2}{M_{Pl}^2}\frac{e^{-k|x-x'|/r_{KK}}}{|\vec x-\vec x'|}
\sim \frac{m_1m_2}{M_{Pl}^{2}}\frac{1}{|\vec x-\vec x'|}
\left(\frac{r_{KK}}{|x-x'|}\right)^n =
\frac{m_1m_2}{m_{KK}^{n}M_{Pl}^{2}}\frac{1}{|x-x'|^{n+1}} =
\frac{m_1m_2}{m_{Pl}^{(n+2)}}\frac{1}{|x-x'|^{n+1}}
\label{grapro}
\ee
where we again replaced the sum over $\vec k$ with the integral.

Therefore, the strength of multidimensional gravity,
i.e. appearance of large $m_{Pl}^{-1}$ instead of the small
$M_{Pl}^{-1}$ can be explained as emergency of many ($\sim k^n$) copies
of the ordinary $4d$ graviton at small
distances $|x-x'| \ll r_{KK}$.

\subsection{Does strong gravity imply strong gravitational
radiation? \label{stgrastra}}

We can now look at the gravitational radiation: it can become
strong if all these copies of the ordinary graviton are
emitted.
Clearly, the only essential change as compared to eq.(\ref{grapro})
is that the role of distance $|x-x'|$ 
is played by the inverse typical frequency of the radiation: the
KK graviton with the mass
$km_{KK}$ looks massless and is emitted with the
same rate as the ordinary graviton only if its frequency
$\omega \gg km_{KK}$
and the relevant $k \sim \omega/m_{KK}$.
This means that the enhancement factor $\sim k^n$ for the gravitational
radiation, converts $M_{Pl}^{-2}$ into
\be
\frac{1}{M_{Pl}^2}\left(\frac{\omega}{m_{KK}}\right)^n =
\frac{1}{m_{Pl}^2}\left(\frac{\omega}{m_{Pl}}\right)^n
\label{grarad}
\ee
Thus, the radiation with the frequencies 
$m_{KK} \ll \omega \ll m_{Pl}$ is not fully multi-dimensional:
it is enhanced as compared to the ordinary $4d$ gravity with
the gravitational coupling $M_{Pl}^{-2}$ but
damped as compared to the TeV-gravity with gravitational
coupling $m_{Pl}^{-2}$.

Due to relativistic effects the radiation frequency $\omega$
can be rather big: for synchrotron radiation of a particle,
moving with the angular velocity $\omega_0 = \frac{v}{R}$
in accelerator ring of radius $R$, the typical
$\omega \sim \gamma^3\omega_0$, where
$\gamma = (1-v^2)^{-1/2} = {\cal E}/m$.
However, the ultrarelativistic particle with high $\gamma$
radiates only inside a narrow cone with the angle
$\vartheta \sim 1/\gamma$, and this means that
only KK gravitons with relatively small $k \sim \vartheta\omega$
are radiated, so that the actual enhancement factor of gravitational
radiation due to the multi-dimensionality in the
ultrarelativistic case gets smaller than (\ref{grarad}):
\be\label{fact}
\frac{1}{m_{Pl}^2}\left(\frac{\omega}{m_{Pl}}\right)^n
\longrightarrow
\frac{1}{m_{Pl}^2}\left(\frac{\vartheta\omega}{m_{Pl}}\right)^n
= \frac{1}{m_{Pl}^2}\left(\frac{{\cal E}}{m_{Pl}}\right)^n
\left(\frac{\vartheta\omega}{{\cal E}}\right)^n
\ee
It is the last factor that causes undesired damping.
If the source particle is charged and kept in an accelerator ring by
magnetic field $B$, then ${\cal E}\omega_0 = eB$ and
\be
\frac{\vartheta\omega}{{\cal E}} \sim
\frac{\gamma^2\omega_0}{{\cal E}}
= \frac{eB\gamma^2}{{\cal E}^2} = \frac{eB}{m^2}
\label{crura}
\ee
i.e. for a given magnetic field
it does not grow with $\gamma$ at all.
To make this ratio of the order of unity, the value of magnetic field
should be critical,
that is, electric field of the same magnitude
$E\sim m^2/e$ would create pairs of particles with mass $m$
at the distances of their Compton wavelength $1/m$.
For electron, the critical field $B\sim 4.4\cdot 10^9$
Tesla and for proton $B\sim 1.5\cdot 10^{16}$ Tesla.

The fields $B \sim 10\ {\rm Tesla}$ used in nowadays accelerators
are so less than the critical field that not only (\ref{crura}) is
terribly small, but $\omega\vartheta$ actually
can not even reach $m_{KK}$ as soon as $n>2$,
and existence of extra dimensions simply does
not affect the synchrotron radiation, see s.\ref{numes}.

\subsection{Intensity of radiation}

The radiated power (radiative energy loss per unit time)
in $4$ space-time dimensions
consists of five different factors:
\be
I \sim \eta g^2\omega^2\vartheta^2 N
\sim \eta_d g_d^2\omega^{d-2}\vartheta^{d-2}
\label{radintens}
\ee

$g$ is a "charge", characterizing coupling of the radiating field
to the source of radiation;

$\omega$ is the typical
frequency of radiated waves;

$\vartheta$ is the typical angle of radiation cone, where emitted
radiation is concentrated; for the ultrarelativistic source
$\vartheta \sim 1/\gamma$, because longitudinal component
of the wave vector (photon momentum) in the forward direction
is Lorentz increased by the $\gamma$-factor, while
transversal components remain intact,
so that the isotropic radiation in the proper frame of the source turns into
a narrow cone around the source velocity; 

$N$ is the number of radiated species, it is restricted to the
number of polarizations in pure $4d$ theories, but in compactified
theories it counts the number of emitted KK particles, and can be
large: $N \sim \left(\frac{\vartheta\omega_c}{m_{KK}}\right)^n$;

$\eta$ is a numerical factor of the order of unity, depending on the
details of radiation process.

\bigskip

The second equality in (\ref{radintens}) provides a description of
the same radiated power $I$ from the $d=4+n$-dimensional point of view:
the factor $N_d$ would count only the number of polarizations
(depending on $d$ and on the spin $s$ of the
radiated field) and is included into $\eta_d$.
Charges in different dimensions are
related in the usual way: $g^{-2} \sim g_d^{-2}r_{KK}^n$,
the numerical factor depending on the shape of the compact dimensions
being also included into $\eta$-factors.

In accordance with (\ref{grarad}) and (\ref{radintens}),
in order to increase the gravitational radiation one could
do the three things: decrease $n$, increase $\omega$ and increase
the radiation of an ordinary graviton, i.e. the coefficient
$g$ of essentially $4d$ origin in front of (\ref{grarad}).
The first possibility is obvious, but depends on Nature
rather than on our effort.
Meanwhile, the other two options,
to raise the radiation frequency and to increase
the probability of $4d$ graviton emission in
realistic experiments, should be considered in more detail.

\subsection{Radiation frequency}

As was already mentioned, the
radiation frequency $\omega$ grows fast with increase
of the energy of ultra-relativistic source.
It is defined by the typical time of changing the field of
the source, $t_{form} = l_{form}/v$.
In the proper frame of the source particle, the formation
length is Lorentz-contracted: $l_{form} \longrightarrow
l_{form}/\gamma$. Furthermore, the detector measures the
Doppler-transformed frequency, differing by the
factor $\big(\gamma(1-\vec n\vec v)\big)^{-1}$ so that
totally for an ultrarelativistic source
\be
\omega \sim \frac{1}{(1-\vec n\vec v)l_{form}} \approx
\frac{1}{\big(\theta^2+\gamma^{-2}\big)l_{form}}.
\label{radfreq}
\ee
where $\theta$ is the angle between the particle velocity
$\vec v$ and the direction $\vec n$ to the observation point.
Dominant is the radiation inside the narrow cone\footnote{
To avoid possible confusion, we mention that
in physical gauge the field itself behaves as
$$\frac{(\sin\theta)^s}{1-v\cos\theta} \sim
\frac{\theta^s}{\theta^2+\gamma^{-2}}$$
and, for gravity ($s=2$), is {\it not} concentrated inside
the cone even for $\gamma \gg 1$;
however, its derivatives and the radiated energy flux are.
}
with $\theta \sim \vartheta=1/\gamma$ and with the typical
frequency of $\omega \sim \gamma^2/l_{form}$.

The formation length $l_{form}$ depends on experimental
setup. For example, for the undulator radiation, produced by
a source moving in an inhomogeneous (periodic)
magnetic field, $l_{form} = l_{ond}$ is the typical
modulation length of the field.
For the synchrotron radiation
$l_{form} = R\vartheta$, since the radiation cone affects a
point-like detector only if emitted from an arc of
angular length $\vartheta$.
It is implicitly assumed in this argument that the radiation
propagates along straight lines and is not affected by
background fields; this is always the case with one
remarkable exception:  
if the source particle is kept on a circle by a
background gravitational field (instead of magnetic one
as in ordinary cyclic accelerators) then the same
background field which curves emitting particle's trajectory
curves the radiation cone as well;
moreover, if $v\approx 1$, then by equivalence principle
the curvatures are exactly the same, and $l_{form}\sim R$.
Another
instructive example is the bremsstrahlung radiation caused by a decelerating force
$F(x)$. Then, there are several possible cases, depending on the deflection angle
$\beta\sim {\vartheta\over m}\int F(x) dx$,
\cite[ref.1]{4drad}. The case of $\beta\gg\vartheta$ is much similar to the
synchrotron radiation, while in the case of small deflection angles the
radiation comes from the whole trajectory, and the formation length is
the length where the acceleration of the particle noticeably varies. E.g.,
for the bremsstrahlung radiation in Coloumb field, the role of formation length is
played by the impact parameter, $b$ so that the radiation spectrum is flat
until the frequency $\gamma^2/b$ and sharply falls off at higher frequencies
\cite[ref.2]{4drad}.
At last, if the scattered particle instantly
changes its speed, the formation length is zero, and there is
no distinguished frequency in the radiation spectrum at all. Moreover, in
this case, the narrow radiation cone is also absent, and there would be no
damping $\vartheta$-factors like those in formula (\ref{fact}), see
\cite{koch}. Unfortunately, it is difficult to instantly stop high
energy particles, hence, another damping would occur due to
the small cross-section of the process.

\subsection{Charges for direct and induced gravitational radiation}

The $4d$ charge $g$ depends on sort of the radiated field.
If we are interested in electromagnetic radiation, then
$g^{em} = e$, moreover, in TeV-gravity models electromagnetic
fields are not allowed to leave the $3$-brane and to propagate
in the bulk, so that $d$-dimensional consideration makes no
sense for them.

For the gravitational radiation, $g = \frac{m}{M_{Pl}} =
\frac{{\cal E}}{\gamma M_{Pl}}$ depends on the mass of the
source particle.
It is in this case that existence of extra dimensions can
lead to a serious enhancement of the radiation, substituting the
tiny $g^{gr}$ by a more reasonable
$g_d^{gr} \sim \frac{m}{m_{Pl}}=\frac{{\cal E}}{\gamma m_{Pl}}$.

Remarkably, the remaining small factor $1/\gamma$ in $g^{gr}_d$
is actually eliminated by an additional phenomenon, known
as induced gravitational radiation \cite{indgrara}.
It is best understood if the radiating particle is charged
and accelerated by a strong electromagnetic field, which we call
background.
The point is that in the presence of a strong background
field photons are mixed with gravitons,
and thus the electromagnetic radiation becomes itself a source
of the gravitational one.
The role of the source for this induced
radiation is played by $\int d^3x BF_{rad}$, where we choose as a background
some magnetic field $B$ to suit accelerator experiments.
In evaluating the effective charge, one should
take into account that the radiated field
$F_{rad}$ is concentrated inside the narrow radiation cone
and falls with the distance from the source particle,
while the background field $B$ is non-vanishing only in
some finite volume.
Detailed calculation \cite{indgrara} shows that
\be
g^{ind} \sim \frac{eBL}{M_{Pl}} =
\frac{{\cal E}}{M_{Pl}}\frac{L}{R}
\ee
where $L$ is the overlap length between the electromagnetic
radiation cone and the background field $B$.
Thus,
\be
\frac{g^{ind}}{g^{gr}} \sim \frac{\gamma L}{R}
\ \ {\rm and} \ \
\frac{I_{ind}}{I_{gr}} \sim \gamma^2\left(\frac{L}{R}\right)^2
\label{ratind}
\ee
Therefore, if $L$ is not much smaller than $R$,
the gravitational radiation from ultrarelativistic source is mostly
induced, while the power of the direct component is damped by additional
factor $1/\gamma^2$.

In fact, existence 
of the induced gravitational
radiation is a universal phenomenon, independent of nature
of the force which accelerates the source.
Be it a background field of any nature, the quanta of {\it this}
field will be emitted, and they will be the source of the induced
radiation.
Neglect of this contribution leads to non-conservation of
the stress tensor and makes the radiation problem badly defined.

It remains to say a few words about the ratio $L/R$.
Normally, the magnetic field $B$ of a cyclotron
is concentrated inside the narrow accelerator
tube of radius $r$, while the synchrotron radiation is tangent
to the ring and goes away from the tube, so that
$L\sim \sqrt{rR}$.
However, one can easily make a dedicated device, with
strong magnetic field along the radiation track,
where $L$ can be made arbitrary large, say, $L\sim R$.
Note that if electrons are used as a source of radiation
(what is important to increase the effect), one also needs
a dedicated device with a strong magnetic field at the exit
of the future linear electron accelerator: first to produce
super-strong electromagnetic synchrotron radiation from
TeV-energy electrons and second to convert it into
induced gravitational radiation.

\subsection{Capture of emitted gravitons}

Even if a relatively strong flux of the gravitational
radiation is produced, it remains to capture it.
This is again somewhat problematic.

Direct measuring of energy losses
is difficult even for the ordinary electromagnetic
synchrotron radiation which is undoubtedly
strong and important for dynamics of particles
in accelerators. Indeed, one can not trace our a single electron, instead
observing the bunches of electrons. Therefore, one can study their movement
only statistically, and the tiny effects of the gravitation radiation can
not be measured in this way on the large background of the electromagnetic
radiation.

The other possibility is to catch the emitted gravitons.
This is, however, also a bad option
because their interaction with any kind of
detectors is damped by tiny ${\cal E}/M_{Pl}$ factors. Indeed,
considering virtual gravitons (e.g., when estimating the two-body
interaction), one sums as many as $\sim M_{Pl}^2$ amplitudes with different KK-graviton
propagators, each one being
damped by the $M_{Pl}^{-2}$ factor, therefore, the total amplitude is not
damped by the Plank mass $M_{Pl}$ at all. This is what we observed in s.2.
On the contrary, when catching gravitons with a detector, i.e.
dealing with the real gravitons when different
amplitudes do not interfere, one should sum the squares of amplitudes,
each proportional to $M_{Pl}^{-4}$, so that the sum of $M_{Pl}^2$ terms now
does not compensate all the Plank factors.

A dedicated M\"ossbauer-style technique
for gravitation wave experiments \cite{chiao},
even if realizable at low frequencies,
can not be applied to ultra-high frequency
synchrotron radiation.
The only option remaining is to use the standard
particle-physics experiments, capturing quantum
particles.
Such search of TeV-gravity gravitons is planned
at LHC \cite{grasearch}. These experiments basically exploit the same two
possibilities we discussed above: either
to look for missed energy in events with hard gravitons emitted, or to search for
events with virtual gravitons using, e.g., their specific angular distribution
because of the spin two of graviton. Unfortunately, there is no way to
exploit the seeming advantages of potentially
strong {\it classical} synchrotron radiation in these
experiments.

\subsection{Numerical estimate and pessimistic conclusion\label{numes}}

In practice, the problem of graviton capturing is
not the main one for the synchrotron radiation of gravitons.
It turns out that the fields
$B \sim 10\ {\rm Tesla} \approx 3\cdot 10^{-15}{\rm GeV}^2$,
used in modern accelerators, are very small: the ratio
$eB/m^2 \sim 10^{-15}$ for protons and $\sim 2\cdot 10^{-9}$
for electrons.
In principle, one could think of using pulses of the magnetic field,
which are already two orders of magnitude higher and can be
further increased with  technology development.
It may, however, be difficult to extend them to large distances
$L$, required to make induced radiation effective.

Still, the main problem is of a more fundamental nature.
Since for $M_{Pl} \sim 10\ {\rm TeV}$ the Kaluza-Klein mass
$m_{Pl}/m_{KK} \sim 10^{30/n}$, the actual number
$\big(\frac{\omega\vartheta}{m_{KK}}\big)^n$ of radiated KK
gravitons is as small as $\sim 10^{30-10n}$ even for electrons,
i.e. for $n\geq 3$ it can not exceed unity, and
multi-dimensionality does not affect radiation at all. For remaining
possibilities of $n\le 2$, see direct Cavendish type experiments in \cite{cav}
and astrophysical bounds in \cite{astro}.
If $n=6$, KK gravitons get produced by electrons only if $B$
exceeds $10^6$ Tesla,
however, the threshold frequency
$\omega \sim m_{KK}/\vartheta = 10^{-5}\gamma m_{Pl} \gg m_{Pl}$
since $\gamma \sim 10^7$ for electrons at $m_{Pl} = 10\ {\rm TeV}$,
and this is unacceptable, because energy loss in
a single radiation act would exceed electron's energy
${\cal E} \sim m_{Pl}$.

Thus, whether TeV-gravity is true or not, it can not lead
to any kind of enhancement of gravitational radiation at LHC
or TeV-energy electronic accelerators.
Strong gravity does not necessarily imply strong
gravitation waves!

\subsection*{Acknowledgements}

We acknowledge conversations with V.Rubakov and T.Tomaras.
Our work is partly supported by the Federal Program of the
Russian Ministry of Industry, Science and Technology No
40.052.1.1.1112, by the grants RFBR
04-02-16538a (Mironov), RFBR 04-02-16880 (Morozov), by the Grant of
Support for the Scientific Schools 8004.2006.2, NWO project
047.011.2004.026, INTAS grant 05-1000008-7865
and ANR-05-BLAN-0029-01 project.

\end{document}